\documentclass[useAMS,usenatbib]{mn2e}
\usepackage{mathrsfs}
\usepackage{graphicx,amssym,color}
\usepackage{threeparttable}

\usepackage{float}
\usepackage{chngcntr}

\newcommand{\beq}{\begin{equation}}
\newcommand{\eeq}{\end{equation}}
\newcommand{\beqa}{\begin{eqnarray}}
\newcommand{\eeqa}{\end{eqnarray}}

\usepackage{color}
\def\gsim { \lower .75ex \hbox{$\sim$} \llap{\raise .27ex \hbox{$>$}} }
\def\lsim { \lower .75ex \hbox{$\sim$} \llap{\raise .27ex \hbox{$<$}} }
\def\proptosim { \lower .75ex \hbox{$\sim$} \llap{\raise .27ex \hbox{$\propto$}} }

\newcommand{\bc}{\begin{center}}
\newcommand{\ec}{\end{center}}

\newcommand{\pc}             {\,{\rm pc}}
\newcommand{\kpc}            {\,{\rm kpc}}
\newcommand{\Mpc}            {\,{\rm Mpc}}
\newcommand{\Msun}           {\,{\rm M}_\odot}

\newcommand{\kms}            {\,{\rm km}\,\,{\rm s}^{-1}}

\title[Dark matter content of MW satellites] {The cold dark matter
  content of Galactic dwarf spheroidals: no cores, no failures, no problem}
\author[A. Fattahi et al. ]{
\parbox[t]{\textwidth}{
       Azadeh Fattahi$^{1}$\thanks{Email: azadehf@uvic.ca}, Julio
         F. Navarro$^{1,2}$, Till Sawala$^{3}$ , Carlos S. Frenk$^{4}$ ,
         \\ Laura V. Sales$^5$, Kyle Oman$^{1}$, Matthieu Schaller$^{4}$ and Jie Wang$^{5}$
}
  \\ \\
\parbox[t]{\textwidth}{ 
$^1$Department of Physics and Astronomy,University of Victoria, PO Box
  3055 STN CSC, Victoria, BC, V8W 3P6, Canada\\
$^2$Senior CIfAR Fellow.\\
$^{3}$Department of Physics, University of Helsinki, Gustaf
  H\"allstr\"omin katu 2a, FI-00014 Helsinki, Finland \\
$^{4}$Institute for Computational Cosmology, Department of Physics,
  University of Durham, South Road, Durham DH1 3LE, United Kingdom\\
$^5$Department of Physics and Astronomy, University of California
  Riverside, 900 University Ave., CA92507, US \\
$^{6}$ Key laboratory for Computational Astrophysics, National
  Astronomical Observatories, Chinese Academy of Sciences, Beijing,
  100012, China \\ } }

\begin{document}

\date{\today}

\pagerange{\pageref{firstpage}--\pageref{lastpage}}
\pubyear{2016}

\maketitle

\label{firstpage}

\begin{abstract}
We examine the dark matter content of satellite galaxies in
 $\Lambda$CDM cosmological hydrodynamical simulations of the Local
 Group from the APOSTLE project. We find excellent agreement between
 simulation results and estimates for the $9$ brightest Galactic
 dwarf spheroidals (dSphs) derived from their stellar velocity
 dispersions and half-light radii. Tidal stripping plays an important
 role by gradually removing dark matter from the outside in,
 affecting in particular fainter satellites and systems of
 larger-than-average size for their luminosity. Our models suggest
 that tides have significantly reduced the dark matter content of
 Can~Ven~I, Sextans, Carina, and Fornax, a prediction that may be
 tested by comparing them with field galaxies of matching luminosity
 {\it and} size. Uncertainties in observational estimates of the dark
 matter content of individual dwarfs have been underestimated in the
 past, at times substantially. We use our improved estimates to revisit
 the `too-big-to-fail' problem highlighted in earlier N-body work. We
 reinforce and extend our previous conclusion that the APOSTLE simulations
 show no sign of this problem. The resolution does {\it
   not} require `cores' in the  dark mass profiles, but, rather,
 relies on revising assumptions and uncertainties in the
 interpretation of observational data and accounting for `baryon
 effects' in the  theoretical modelling.
\end{abstract}


\section{Introduction}
\label{SecIntro}

The steep slope of the dark matter halo mass function at the low-mass
end is a defining characteristic of the $\Lambda$CDM
cosmological paradigm. It is much steeper than the faint-end slope of
the galaxy stellar mass function, implying that low-mass CDM haloes are
significantly more abundant than faint dwarf galaxies
\citep{Moore1999b,Klypin1999}. This discrepancy is usually reconciled
by assuming that dwarfs form preferentially in relatively massive
haloes, because cosmic reionization and the energetic
feedback from stellar evolution are effective at removing baryons from
the shallow gravitational potential of low-mass systems and at
curtailing their star forming activity
\citep{Bullock2000,Benson2002,Somerville2002}.

Such scenario makes clear predictions for the stellar mass -- halo
mass relation at the faint end. A simple -- but powerful and widely
used -- parameterization of this prediction is obtained from
abundance matching (AM) modeling, where galaxies and CDM haloes are
ranked by mass and matched to each other respecting their relative
ranked order \citep[see,
e.g.,][]{Frenk1988,Vale2004,Guo2011,Moster2013,Behroozi2013}. Halo masses may thus be
derived from stellar masses, yielding clear predictions amenable to
observational testing.

Most such tests rely on using kinematic tracers such as rotation
speeds or velocity dispersions to estimate the total gravitational
mass enclosed within the luminous radius of a galaxy. Its dark matter
content, computed after subtracting the contribution of the baryons,
may then be used to estimate the total virial\footnote {We define
  virial quantities as those corresponding to the radius where the
  spherical mean density equals 200 times the critical density for
  closure, $3H^2/8\pi G$. Virial quantities are identified by a
  ``200'' subscript.} mass of the system. Such estimates rely heavily
on the similarity of the mass profiles of CDM haloes \citep[][referred
to hereafter as NFW]{Navarro1996,Navarro1997}, and involve a fairly
large extrapolation, since virial radii are typically much larger than
galaxy radii.

These tests have revealed some tension between the predictions of AM
models and observations. \citet{Boylan-Kolchin2011}, for example,
estimated masses for the most luminous Galactic satellites that were
lower than those of the most massive substructure haloes in N-body
simulations of Milky Way-sized haloes from the Aquarius Project
\citep{Springel2008b}. \citet{Ferrero2012} reported a related finding
when analyzing the dark matter content of faint dwarf irregular
galaxies in the field: many of them implied total virial masses well
below those predicted by AM models. Subsequent work has highlighted
similar results both in the analysis of M31 satellites
\citep{Tollerud2014,Collins2014}, as well as in other samples of field
galaxies \citep{Garrison-Kimmel2014,Papastergis2015}.

These discrepancies may in principle be reconciled with $\Lambda$CDM
in a number of ways.  One possibility is to reconsider virial mass
estimates based on the dark mass enclosed by the galaxy, a procedure
that is highly sensitive to assumptions about the halo mass profile. A
popular revision assumes that the assembly of the galaxy may lead to a
reshuffling of the mass profile, pushing dark matter out of the inner
regions and creating a constant-density `core' in an otherwise cuspy
NFW halo
\citep[e.g.,][]{Navarro1996b,Mashchenko2006,Governato2012}. 

These cores allow dwarf galaxies to inhabit massive haloes despite
their relatively low inner dark matter content. This option has received
some support from hydrodynamical simulations \citep[see, e.g.,][for a
review]{Pontzen2014} although the results are sensitive to how star
formation and feedback are implemented. Indeed, no consensus has yet
been reached over the magnitude of the effect, its dependence on mass,
or even whether such cores exist at all \citep[see, e.g.,][and
references
therein]{Parry2012,Garrison-Kimmel2013,DiCintio2014,Schaller2015a,Oman2015, Onorbe2015}.

A second possibility is that Galactic satellites have been affected by
tidal stripping, which would preferentially remove dark matter
\citep[e.g.,][]{Penarrubia2008b} and, therefore, act to reduce their
dark mass content, much as the baryon-induced `cores' discussed in
the preceding paragraph. This proposal would not help to solve the
issue raised by field dwarf irregulars \citep{Ferrero2012} nor the low
dark matter content of Galactic satellites (tides are, of course,
already included in N-body halo simulations), unless baryon-induced
cores help to enhance the effects of tides, as proposed by
\citet{Zolotov2012} and \citet{Brooks2014}.

A third option is to revise the abundance matching prescription so as
to allow dwarf galaxies to inhabit haloes of lower mass. This would be
the case if some galaxies simply fail to form (or are too faint to
feature in current surveys) in haloes below some mass: once these
``dark'' systems are taken into account, the AM stellar mass -- halo
mass relation would shift to systematically lower virial masses for
given stellar mass, as pointed out by \citet{Sawala2013}.

The existence of `dark' subhaloes does not, on its own, solve the
problem pointed out by \citet{Boylan-Kolchin2011}, which is usually
referred to as the `too-big-to-fail' problem \citep[hereafter TBTF,
see also][]{Boylan-Kolchin2012}. Indeed, associating dwarfs with lower
halo masses would not explain why many of the most massive
substructures in the Aquarius haloes seem inconsistent with the
kinematic constraints of the known Galactic satellites.

One explanation might be that fewer massive subhaloes are present in
the Milky Way (MW) than in Aquarius haloes. Since the number of
substructures scales with the virial mass of the main halo, a lower
Milky Way mass would reduce the number of massive substructures, thus
alleviating the problem
\citep{Wang2012a,Vera-Ciro2013,Cautun2014}. Another possibility is
that dark-matter-only (DMO) simulations like Aquarius overestimate the
subhalo mass function. Low mass haloes are expected to lose most of
their baryons to cosmic reionization and feedback, a loss that stunts
their growth and reduces their final mass. The effect is limited in
terms of mass (baryons, after all, make up only $17$ per cent of the total
mass of a halo) but it can have disproportionate consequences on the number
of massive substructures given the steepness of the subhalo mass
function \citep{Guo2015,Sawala2016}.

\begin{figure}
  \hspace{-0.2cm}
  \resizebox{8cm}{!}{\includegraphics{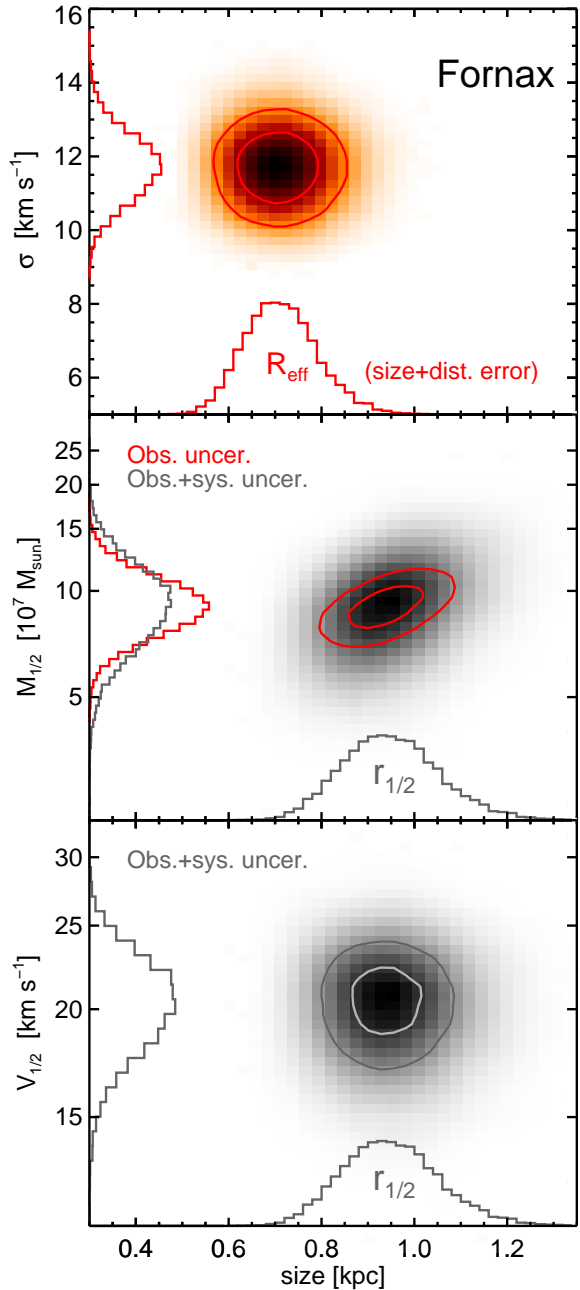}}\\%
  \caption{{\it Top}: Stellar velocity dispersion and effective radius
    ($R_{\rm eff}$) of the Fornax dSph. The $R_{\rm eff}$ distribution
    is obtained by convolving uncertainties in distance and in the
    observed angular half-light radius, using uncertainties from the
    literature and assuming Gaussian error distributions. {\it
      Middle}: Dynamical mass within the deprojected 3D half-light
    radius ($r_{1/2}$) of Fornax, calculated using eq.~\ref{eqW10}
    \citep{Wolf2010}. The red histogram shows the result of
    propagating the observational uncertainties, whereas the grey
    histogram adds a $23$ per cent base modeling uncertainty, as
    suggested by \citet{Campbell2016}. {\it Bottom}: Circular velocity
    at $r_{1/2}$ ($V_{1/2}$), including both observational and
    systematic uncertainties, calculated from the final $M_{1/2}$
    distribution (middle panel). Unlike $M_{1/2}$, $V_{1/2}$ is
    independent of $r_{1/2}$. Contours in all panels enclose $50$ per
    cent and $80$ per cent of the distributions. }
\label{FigError}
\end{figure}

\begin{figure}
  \hspace{-0.2cm}
  \resizebox{8cm}{!}{\includegraphics{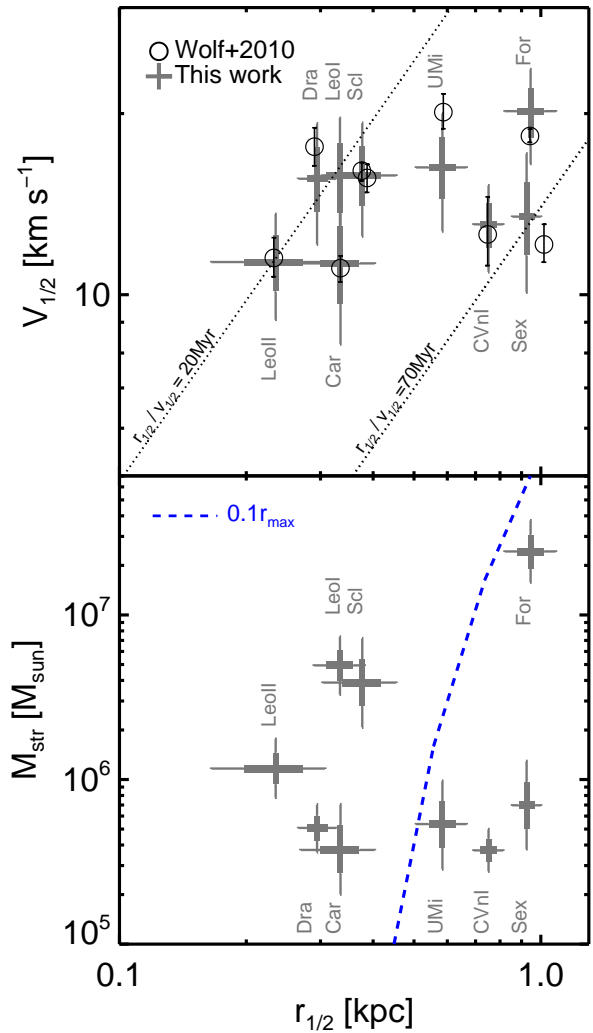}}\\%
  \caption{{\it Top:} Circular velocity at the half-light radius of
    Milky Way classical dSphs. Open circles show the results of
    \citet{Wolf2010} with $1\sigma$ error. The bar-and-whisker symbols
    show the results of this work, including observational and
    systematic uncertainties (see, e.g., the bottom panel of
    Fig.~\ref{FigError} for the case of the Fornax dSph). The thick
    and thin portions illustrate interquartile and $10$--$90^{\rm th}$
    percentile intervals, respectively. Our results suggest that $V_{1/2}$
    uncertainties have been underestimated in previous work. Slanted
    lines show objects with constant crossing time, as labelled. {\it
      Bottom:} Stellar mass derived for the $9$ Galactic dSphs, shown
    as a function of their half-light radius. The blue dashed line
    indicates the characteristic halo mass -- radius dependence of
    APOSTLE centrals, computed from the fit shown in
    Fig.~\ref{FigMstarM200}. The line divides the sample in two groups
    of compact objects resilient to tides and more extended systems
    where tidal effects may be more apparent.}
  \label{FigVR}
\end{figure}

We explore these issues here using $\Lambda$CDM cosmological
hydrodynamical simulations of the Local Group from the
APOSTLE\footnote{A Project Of Simulating The Local Environment}
project \citep{Fattahi2016,Sawala2016}. These simulations use the same code as
the EAGLE project, whose numerical parameters have been calibrated to
reproduce the galaxy stellar mass function and the distribution of
galaxy sizes \citep{Schaye2015,Crain2015}. Our analysis complements
that of \citet{Sawala2016}, who showed that APOSTLE reproduces the
Galactic satellite luminosity/stellar mass function, as well as the
total number of galaxies brighter than $10^5 \Msun$ within the Local
Group.

We extend here the TBTF discussion of that paper by reviewing the
accuracy of observational constraints (Sec.~\ref{SecObs}), which are
based primarily on measurements of line-of-sight velocity dispersions
and the stellar half-mass radii ($r_{1/2}$) of `classical' (i.e.,
$M_V<-8$) Galactic dwarf spheroidals (dSphs), and by focusing our
analysis on the actual mass enclosed within $r_{1/2}$ rather than on
extrapolated quantitites such as the maximum circular velocity of
their surrounding haloes. We also highlight the effect of Galactic
tides, and identify the satellites where such effects might be more
easily detectable observationally.

This paper is organized as follows. We begin by reviewing in
Sec.~\ref{SecObs} the observational constraints on the mass of
Galactic dSphs. We then describe briefly our simulations and discuss
our main results in Sec.~\ref{SecResults}, and conclude with a summary
of our main conclusions in Sec.~\ref{SecConc}.

\section{The mass of Milky-Way dwarf spheroidals}
\label{SecObs}

Dwarf spheroidals (dSph) are dispersion-supported stellar systems,
with little or no gaseous content. Their stellar velocity dispersion
may be combined with the half-mass radius, $r_{1/2}$, to estimate the
total mass enclosed within $r_{1/2}$. This estimate depends only
weakly on the velocity anisotropy, provided that the system is in
equilibrium, close to spherically symmetric, and that its velocity
dispersion is relatively flat \citep{Walker2009d,Wolf2010}. In that
case, the latter authors show that the total mass enclosed within the
(deprojected) half-mass radius is well approximated by
\begin{equation}
 M_{1/2}=3\,G^{-1}\,\sigma_{\rm los}^2\,r_{1/2},
\label{eqW10}\end{equation} 
where $\sigma_{\rm los}$ is the luminosity-weighted line-of-sight velocity
dispersion of the stars and $r_{1/2}$ has been estimated from the
(projected) effective radius, $R_{\rm eff}$, using $r_{1/2}=(4/3)R_{\rm
  eff}$. 

The velocity dispersion profiles of the Milky Way classical dSph satellites are
indeed nearly flat \citep{Walker2007,Walker2009d}, and eq.~\ref{eqW10}
has been used to estimate $M_{1/2}$ or, equivalently, the circular
velocity at $r_{1/2}$, $V_{1/2}=\sqrt{GM_{1/2}/r_{1/2}}$, for many of them.  The two
parameters needed for eq.~\ref{eqW10} are inferred from (i) individual
stellar velocities; (ii) the angular projected half-light radius; and
(iii) the distance modulus, each of which is subject to observational
uncertainty. A lower limit on the uncertainty in $M_{1/2}$ may thus be
derived by propagating the uncertainties in each of those three
quantities.  We shall adopt the most up-to-date values from the
\citet{McConnachie2012} Local Group compilation\footnote{http://www.astro.uvic.ca/$\sim$alan/Nearby\_Dwarf\_Database.html}
as the main source of observational data. Table \ref{TabSat} lists our
adopted values for the $9$ dSphs within $300$ kpc from the Milky
Way. (We have excluded the Sagittarius dwarf from our analysis because
it is in the process of being tidally disrupted.)
 
We show in Fig.~\ref{FigError} the error budget (assumed Gaussian
unless otherwise specified) for the case of the Fornax dSph, one of
the best studied Galactic dSphs. The top panel illustrates the errors
in $\sigma_{\rm los}$ and $R_{\rm eff}$, including errors in the distance
and the angular half size. The red histogram in the middle panel of
Fig.~\ref{FigError} shows the result of applying eq.~\ref{eqW10}, after
transforming $R_{\rm eff}$ into 3D $r_{1/2}$, assuming no
additional error.

\begin{figure*}
  \bc \hspace{-0.2cm}
  \resizebox{17cm}{!}{\includegraphics{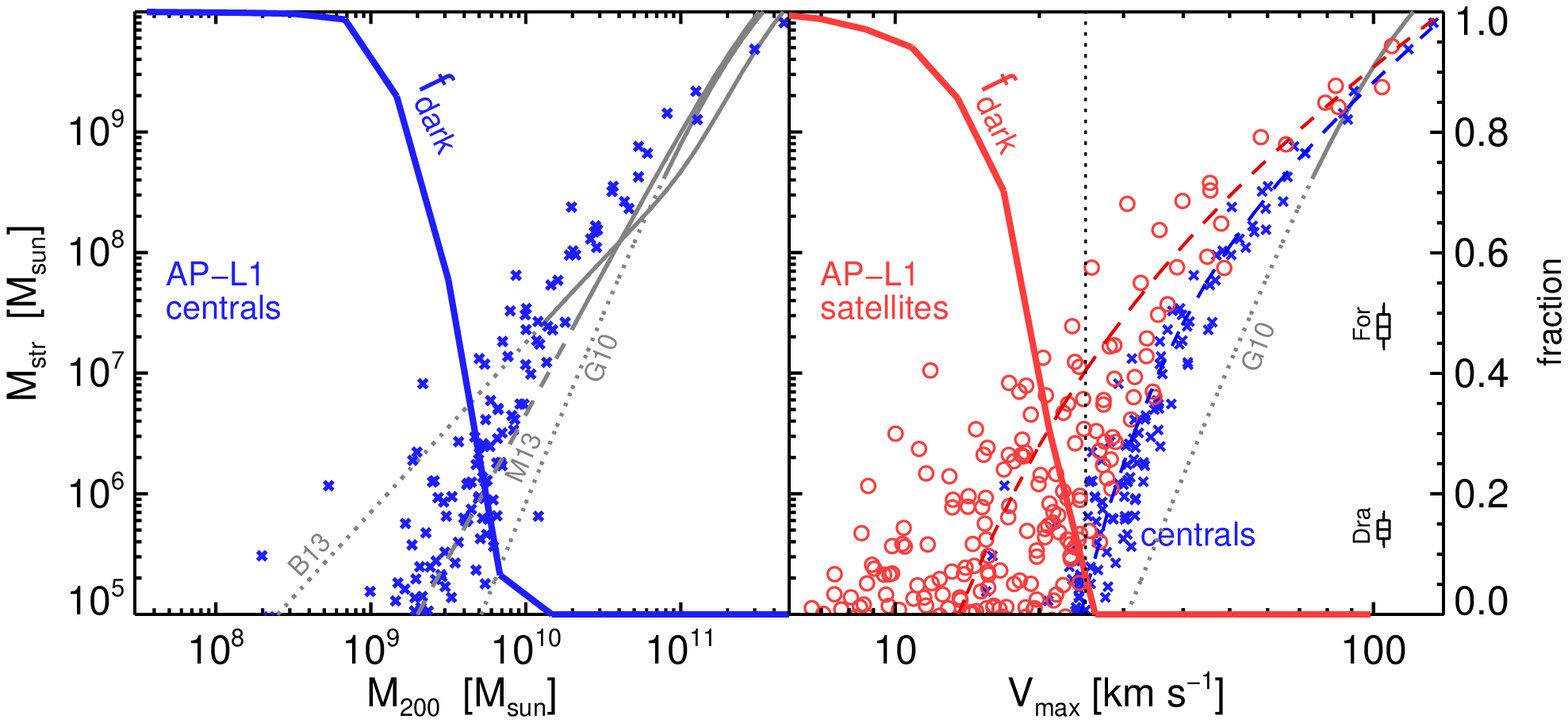}}\\%
  \caption{{\it Left}: Stellar mass -- halo mass relation for
    `central' galaxies in the highest resolution APOSTLE runs
    (L1). The abundance matching relations of \citet{Guo2010},
    \citet{Moster2013} and \citet{Behroozi2013} are shown for
    reference, labelled as G10, M13, and B13, respectively. The dotted
    portion of these curves indicates extrapolation of their formulae
    to low masses.  The fraction of `dark' systems in APOSTLE (i.e.,
    no star particles) as a function of virial mass is indicated by
    the curve labelled `$f_{\rm dark}$', with the scale shown on the
    right axis. {\it Right:} Stellar mass versus maximum circular
    velocity ($V_{\rm max}$) of centrals and satellite galaxies (at
    $z=0$ for both) in APOSTLE, shown as blue crosses and red circles,
    respectively. The offset between field and satellite galaxies is
    due to loss of mass, mostly dark matter, caused by tidal
    stripping. The fraction of `dark' subhaloes is shown by the solid
    red curve. There are no dark subhaloes with $V_{\rm max}>25\kms$.
    Blue and red dashed lines are fits to the central and satellite
    stellar mass -- $V_{\rm max}$ relations, respectively, of the form
    $M_{\rm str}/\Msun=M_0\,\nu^{\alpha}\exp(-\nu^{\gamma})$, where
    $\nu$ is the velocity in units of $50 \kms$. Best fits have
    $(M_0,\alpha,\gamma)$ equal to $(3.0\times10^8, 3.36, -2.4)$ and
    $(8.0\times10^8,2.70,-1.3)$ for centrals and satellites,
    respectively. For illustration, we indicate the stellar mass of
    Fornax and Draco with box-and-whisker symbols, at an arbitrary
    value of $V_{\rm max}$.}
\label{FigMstarM200} \ec
\end{figure*}

The error propagation results in a $30$ per cent uncertainty in $M_{1/2}$,
with some covariance with that in $r_{1/2}$. Note that this
uncertainty is substantially larger than the $\sim 7$ per cent uncertainty
quoted for Fornax by \citet{Wolf2010}. Furthermore, the uncertainty
shown by the red histogram in Fig.~\ref{FigError} assumes that
applying eq.~\ref{eqW10} introduces no additional error. This
assumption has been recently examined by \citet{Campbell2016}, who
conclude that such modeling has a base systematic uncertainty of
$\sim 23$ per cent, even when half-mass radii and velocity dispersions are known
with exquisite accuracy. We therefore add this in quadrature to obtain
the grey histogram in the middle panel of
Fig.~\ref{FigError}. Finally, using the circular velocity, $V_{1/2}$,
instead of $M_{1/2}$ removes the covariance between mass and radius
(see bottom panel of the same figure), so we shall hereafter adopt
$V_{1/2}$ for our analysis.

We have followed this procedure to compute $r_{1/2}$ and $V_{1/2}$ for
all $9$ classical MW dSphs, and quote their values and uncertainties
in Table~\ref{TabSat}.  Note that in a number of cases these
uncertainties are well in excess of those assumed in recent work. This
may also be seen in the top panel of Fig. \ref{FigVR}, where the grey
crosses indicate our results and compare them with the values quoted
by \citet{Wolf2010}, shown by the open circles. Some of the
differences may be ascribed to revisions to the observational data
from more recent studies and some to the increase in the error due to
the base systematic uncertainty discussed above.

We have also estimated stellar masses for all Galactic dSphs in order
to facilitate comparison with simulated data. We do this by using the
$V$-band magnitude and distance modulus from the compilation of
\citet{McConnachie2012}, and stellar mass-to-light ratios from
\citet{Woo2008}. Errors in $V$-band magnitude and distance modulus are
taken from \citet{McConnachie2012}. \citet{Woo2008} do not provide
uncertainties in the mass-to-light ratios, so we assume a constant
$10$ per cent uncertainty for all systems. We list all observable quantities
and derived stellar masses in Table \ref{TabSat} and show, for future
reference, the relation between stellar mass and half-mass radius in
the bottom panel of Fig.~\ref{FigVR}.

\section{Results}
\label{SecResults}

\subsection{The Local Group APOSTLE simulations}
\label{SecSim}

We shall use results from the APOSTLE project, a suite of
cosmological hydrodynamical simulations of $12$ independent volumes chosen to resemble
the Local Group of Galaxies (LG), with a relatively isolated dominant
pair of luminous galaxies analogous to M31 and the Milky Way. A full description
of the volume selection procedure and of the simulations is presented in
\citet{Fattahi2016} and \citet{Sawala2016}. We briefly summarize here
the main parameters of the simulations relevant to our analysis.

LG candidate volumes for resimulation were selected from a
dark-matter-only (DMO) simulation of a $(100\Mpc)^3$ cosmological box with
$1620^3$ particles \citep[known as DOVE,][]{Jenkins2013}.  DOVE adopts cosmological
parameters consistent with 7-year {\it Wilkinson Microwave Anisotropy
  Probe} \citep[WMAP-7,][]{Komatsu2011} measurements, as follows:
$\Omega_m=0.272$, $\Omega_{\Lambda}=0.728$, $h=0.704$,
$\sigma_8=0.81$, $n_s=0.967$.
 
Each APOSTLE volume includes a relatively-isolated pair of haloes with
kinematics consistent with the MW--M31 pair; in particular: (i) the
pair members are separated by $600$ to $1000 \kpc$; (ii) they are
approaching each other with velocities in the range $(-250,0)\kms$;
and (iii) their relative tangential velocities do not exceed $100
\kms$. The virial mass of the pair members are in the range $(5\times
10^{11},2.5\times 10^{12}) \Msun$, and the combined virial masses are
in the range $(1.6\times 10^{12},4\times 10^{12}) \Msun$. An isolation
criterion is also adopted to ensure that no halo more massive than the
smaller of the pair is found within $2.5 \Mpc$ from the pair
barycentre.

APOSTLE volumes were resimulated using the code developed for the
EAGLE simulation project \citep{Schaye2015, Crain2015}. The code is a
highly modified version of the Tree-PM/SPH code, P-Gadget3
\citep{Springel2005b,Schaller2015c}, with subgrid implementations for
star formation, radiative cooling, metal enrichment, uniform UV and
X-ray background (cosmic reionization), feedback from evolving stars,
as well as the formation and growth of supermassive black holes and
related feedback. APOSTLE runs use the parameters of the `Ref' model
described in \citet{Schaye2015}. The EAGLE galaxy formation model has been
calibrated to reproduce the galaxy stellar mass function and sizes in
the mass range $10^8$--$10^{11}\Msun$ at $z=0.1$. This leads to
relatively `flat' rotation curves for luminous galaxies that agree
well with observations \citep{Schaller2015a}.

\begin{figure*}
  \bc \hspace{-0.2cm}
\resizebox{17cm}{!}{\includegraphics{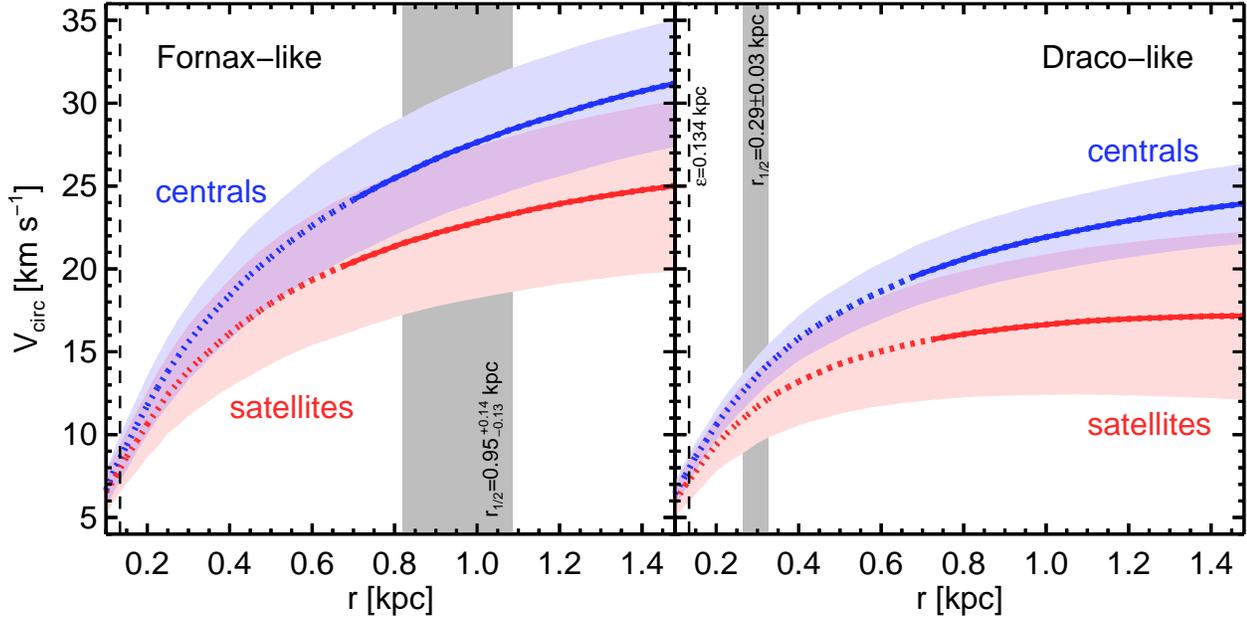}}\\%
\caption{Circular velocity profiles of Fornax- and Draco-like
  satellites (red) and centrals (blue), as labelled. The solid lines
  indicate the average profiles, which turn to dotted inside the
  \citet{Power2003} convergence radius. Shaded regions indicate
  $\pm1\sigma$ deviations. The grey vertical bars bracket the $10^{\rm th}$ and
  $90^{\rm th}$ percentiles of the half-light radii of Fornax and Draco,
  respectively. Note that, although both sets of satellites have been
  heavily stripped, the dark matter content within $r_{1/2}$ has been
  more significantly affected in the case of Fornax, given its
  relatively large size.}
\label{FigTS} \ec
\end{figure*}

The APOSTLE project aims to simulate each volume at three different
resolution levels (L1 to L3). At the time of writing, all $12$ APOSTLE
volumes (AP-$1$ to AP-$12$) have been resimulated at L3 and L2
resolution levels with gas particle mass of $\sim 10^6\Msun$ and
$\sim10^{5}\Msun$, respectively. Three volumes \citep[AP-1, AP-4,
  AP-11, see][]{Fattahi2016} have also been completed at the highest
resolution level, L1, with gas particle mass of $\sim 10^4$, DM
particle mass of $\sim5\times 10^4 \Msun$, and maximum gravitational
softening of $134\pc$. In this paper, we shall use mainly results from
the APOSTLE L1 runs, unless otherwise specified.

Dark matter haloes in APOSTLE are identified using a
friends-of-friends \citep[FoF,][]{Davis1985} algorithm with linking
length equal to $0.2$ times the mean interparticle separation. The FOF
algorithm is run on the dark matter particles; gas and star particles
acquire the FoF membership of their nearest DM particle. Self-bound
substructures inside each FoF halo are then found recursively using
the SUBFIND algorithm \citep{Springel2001a,Dolag2009}. We will
hereafter refer to the main structure of each FoF halo as its
`central', and to the self-bound substructures as its `satellites'. MW
and M31 analogues in the simulations are referred to as primary
galaxies.

\subsection{Stellar mass -- halo mass relation in APOSTLE }
\label{SecMstarM200}

Abundance matching models provide the relation between the stellar
mass and virial mass of galaxies by assuming that every dark matter
halo hosts a galaxy and that there is a monotonic correspondence
between stellar mass and halo mass.  The relation is best specified in
the regime where the galaxy stellar mass function is well determined
($M_{\rm str} > 10^7 \Msun$), but is routinely extrapolated to lower
masses, usually assuming a power-law behaviour \citep[][hereafter G10,
B13, and M13, respectively]{Guo2010,Behroozi2013,Moster2013}.

We compare the APOSTLE stellar mass -- halo mass relation with the
predictions of three different AM models in the left panel of
Fig. \ref{FigMstarM200}. Stellar masses, $M_{\rm str}$, are measured
for simulated galaxies within the `galactic radius', $r_{\rm gal}$,
defined as $0.15$ times the virial radius the halo. This radius
contains most of the stars and cold, star-forming gas of the main
(`central') galaxy of each FoF halo. When considering galaxies
inhabiting subhaloes (`satellites'), whose virial radii are not well
defined, we shall compute $r_{\rm gal}$ using their maximum circular
velocity, $V_{\rm max}$, after calibrating the $V_{\rm max}$--$r_{\rm
  gal}$ relation\footnote{Specifically, we used $r_{\rm
    gal}/$kpc$=0.169\,(V_{\rm max}/\kms)^{1.01}$} of the centrals.

\begin{figure*}
  \hspace{-0.2cm}
  \resizebox{17cm}{!}{\includegraphics{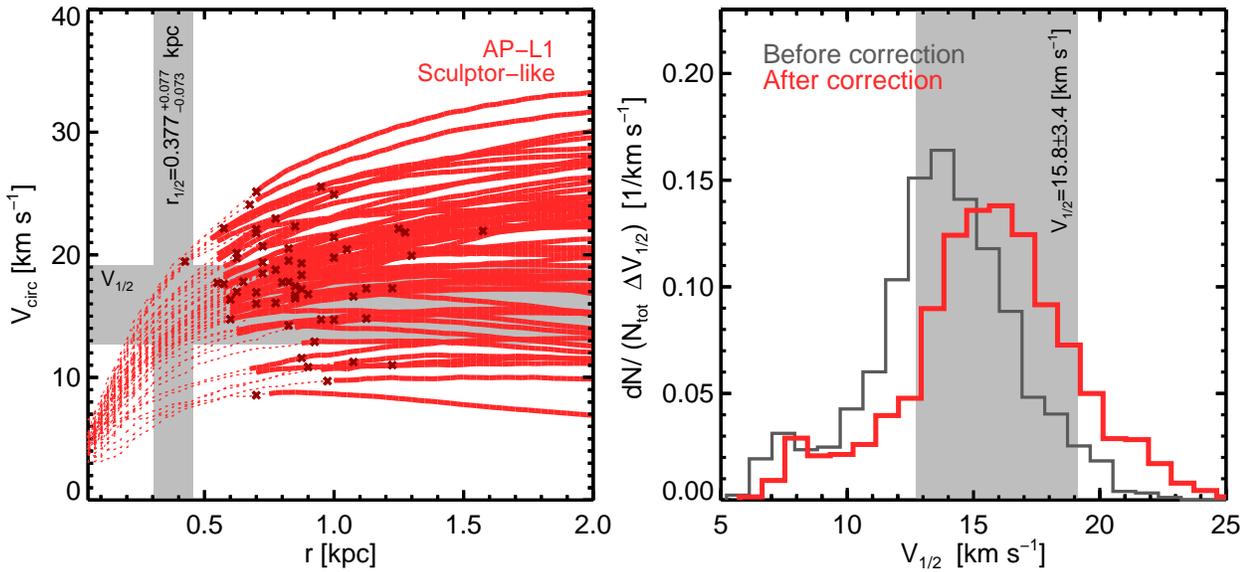}}\\%
  \caption{{\it Left}: Circular velocity curves of Sculptor-like
    satellites in APOSTLE, i.e., systems within $300$ kpc from either
    of the main primaries with stellar mass matching that of Sculptor
    within $3\sigma$ of the central value given in
    Table~\ref{TabSat}. Curves turn from solid to dotted inside the
    \citet{Power2003} convergence radius (for $\kappa=0.6$, see
    Appendix~\ref{SecAppendix}). The grey vertical band indicates the
    half-light radius of Sculptor and the corresponding $10$--$90^{\rm
      th}$ precentile interval. Small crosses indicate the half-mass
    radii of Sculptor-like simulated satellites. {\it Right}:
    Distribution of circular velocities of Sculptor-like satellites,
    measured at the half-light radius of the Sculptor dSph (grey band
    on left). The red histogram shows the distribution after applying
    the resolution correction described in
    Appendix~\ref{SecAppendix}. The grey vertical band corresponds to
    $V_{1/2}$ of Scultpr ($10$--$90^{\rm th}$ precentile interval).}
\label{FigSim}
\end{figure*}

The left panel of Fig.~\ref{FigMstarM200} shows that APOSTLE centrals
do not form `stochastically' in low mass haloes as envisioned in some
models \citep[e.g.,][]{Guo2015}, but, rather, follow a tight
stellar mass -- halo mass relation that deviates systematically from the AM
predictions/extrapolations of G10 and M13.  APOSTLE galaxies of given
stellar mass live in haloes systematically less massive than
extrapolated by those models but more massive than the B13
extrapolation. This reflects the fact that the galaxy stellar mass
function of faint galaxies is rather poorly known, and that AM
`predictions' must be considered with care in this mass regime.

The systematic offset between the G10 and M13 AM extrapolations and
our numerical results has been discussed by
\citet{Sawala2013,Sawala2015}, who trace the disagreement at least in
part to the increasing prevalence of `dark'\footnote{These are
  systems with no stars in APOSTLE L1, or, more precisely,
  $M_{\rm str}<10^4\Msun$, the mass of a single baryonic particle.}
haloes with decreasing virial mass. The effect of these dark systems
is not subtle, as shown by the thick solid blue line in
Fig. \ref{FigMstarM200}. This indicates the fraction of APOSTLE haloes
that are dark (scale on right axis); only {\it half} of
$10^{9.5}\Msun$ haloes harbor luminous galaxies in APOSTLE. The
`dark' fraction increases steeply with decreasing mass: $9$ out of
$10$ haloes with $M_{200}=10^9\Msun$ are dark, and fewer than $1$ in
$50$ haloes with virial mass $\sim 10^{8.8}\Msun$ are luminous.

One might fear that the deviation from the AM prediction shown in
Fig.~\ref{FigMstarM200} might lead to a surplus of faint galaxies in
the Local Group. This is not the case; as discussed by
\citet{Sawala2016}, APOSTLE volumes contain $\sim 100$ galaxies with
$M_{\rm str}>10^5\Msun$ within $2$ Mpc from the LG barycentre, only
slightly above the $\sim 60$ known such galaxies in the compilation of
\citet{McConnachie2012}, which might still be incomplete due to the
difficulty of finding dwarf galaxies in the Galactic `zone of
avoidance'. We shall hereafter adopt $10^5\Msun$ (which corresponds
roughly to a magnitude limit of $M_V\sim -8$) as the minimum galaxy
stellar mass we shall consider in our discussion. In APOSTLE L1 runs
these systems inhabit haloes of $M_{200}\sim 2\times 10^9\Msun$ (three
quarters of which are `dark'), and are resolved with a few tens of
thousands of particles.

\subsection{Tidal stripping effects} 
\label{SecTS}

The right-hand panel of Fig.~\ref{FigMstarM200} is analogous to the
left but using $V_{\rm max}$ (at $z=0$) as a measure of mass
\citep[see also][]{Sales2016a}. This allows the satellites in APOSTLE
(open circles) to be included and compared with centrals (blue
crosses). Satellites clearly deviate from centrals and push the offset
from the G10 abundance matching prediction even further. This is
mainly the result of tidal stripping, which affects disproportionately
the dark matter content of a galaxy, reducing its $V_{\rm max}$ and
increasing its scatter at a given stellar mass \citep[see, e.g.,][and
references therein]{Penarrubia2008b}.

Despite the large scatter, a few results seem robust. One is that
{\it every} subhalo with $V_{\rm max}>25 \kms$ is host to a satellite
more massive than $10^5 \Msun$. This implies that the number of
massive subhaloes provides a firm lower limit to the total number of
satellites at least as bright as the `classical' dSphs, an issue to
which we shall return below.

A second point to note is that the effects of tidal stripping increase
with decreasing stellar mass. Indeed, the $V_{\rm max}$ of
Fornax-like\footnote{We match Galactic satellites with APOSTLE dwarfs
  by stellar mass. For example, we refer to systems as Fornax-like if
  their $M_{\rm str}$ match Fornax's within $3\sigma$. Fornax-like
  {\it satellites} are those within $300\kpc$ of any of the APOSTLE
  primaries; Fornax-like {\it centrals} refer to field galaxies beyond
  that radius.}  centrals is, on average, only $37$ per cent higher
than that of corresponding satellites; the difference, on the other
hand, increases to $67$ per cent in the case of Draco. This trend
arises because dynamical friction erodes the orbits of massive
satellites much faster than those of less luminous systems, leading to
rapid merging or full disruption. As a result, surviving luminous
satellites have, on average, been accreted more recently and have been
less stripped than fainter systems \citep[see, e.g.,][]{Barber2014}.

This does not necessarily imply that the {\it stellar} components of
fainter satellites have been more affected by stripping. Tides are more
effective at removing (mostly dark) mass from the outskirts of a
subhalo than from the inner regions, so their effects on the stellar
component (for a given orbit) will be sensitive to the size of the
satellite. This may be appreciated from Fig.~\ref{FigTS}, where we
show the average circular velocity profiles of both Fornax- and
Draco-like satellites and centrals. The outer regions are clearly more
heavily stripped, implying that satellites that are physically large
for their luminosity should show clearer signs of stripping than their
more compact counterparts.

In other words, dSphs like Can Ven~I or Sextans, for example, are much
more likely to have been affected by tides than Draco or
Leo~II. Fig.~\ref{FigVR} illustrates this in two different ways. In
the top panel, the latter are seen to have much shorter crossing times
than the former, making them more resilient to tides. Similarly, in
the bottom panel, the former are shown to be physically larger than
the latter both at fixed stellar mass and in terms of the
characteristic radius of their host haloes (according to the stellar
mass -- halo mass relation for APOSTLE centrals shown in
Fig.~\ref{FigMstarM200}; see blue dashed line).

Thus, although our results suggest that satellites and field galaxies
of similar $M_{\rm str}$ are expected to inhabit haloes of different
$V_{\rm max}$, the difference might not translate directly into an
observable deficit in their dark matter content\footnote{Indeed,
  \citet{Kirby2014} argue that no large differences seem to exist
  between field and satellite galaxies in the LG.}. This is because
$V_{\rm max}$ is usually reached at radii much larger than the stellar
half-mass radii where kinematic data provide meaningful
constraints. Given the large scatter in $V_{\rm max}$ at a given
stellar mass shown by APOSTLE satellites, it is important to compare
simulations and observations {\it at the same radii}. We explore this
next.

\begin{figure*}
  \hspace{-0.2cm}
  \resizebox{17cm}{!}{\includegraphics{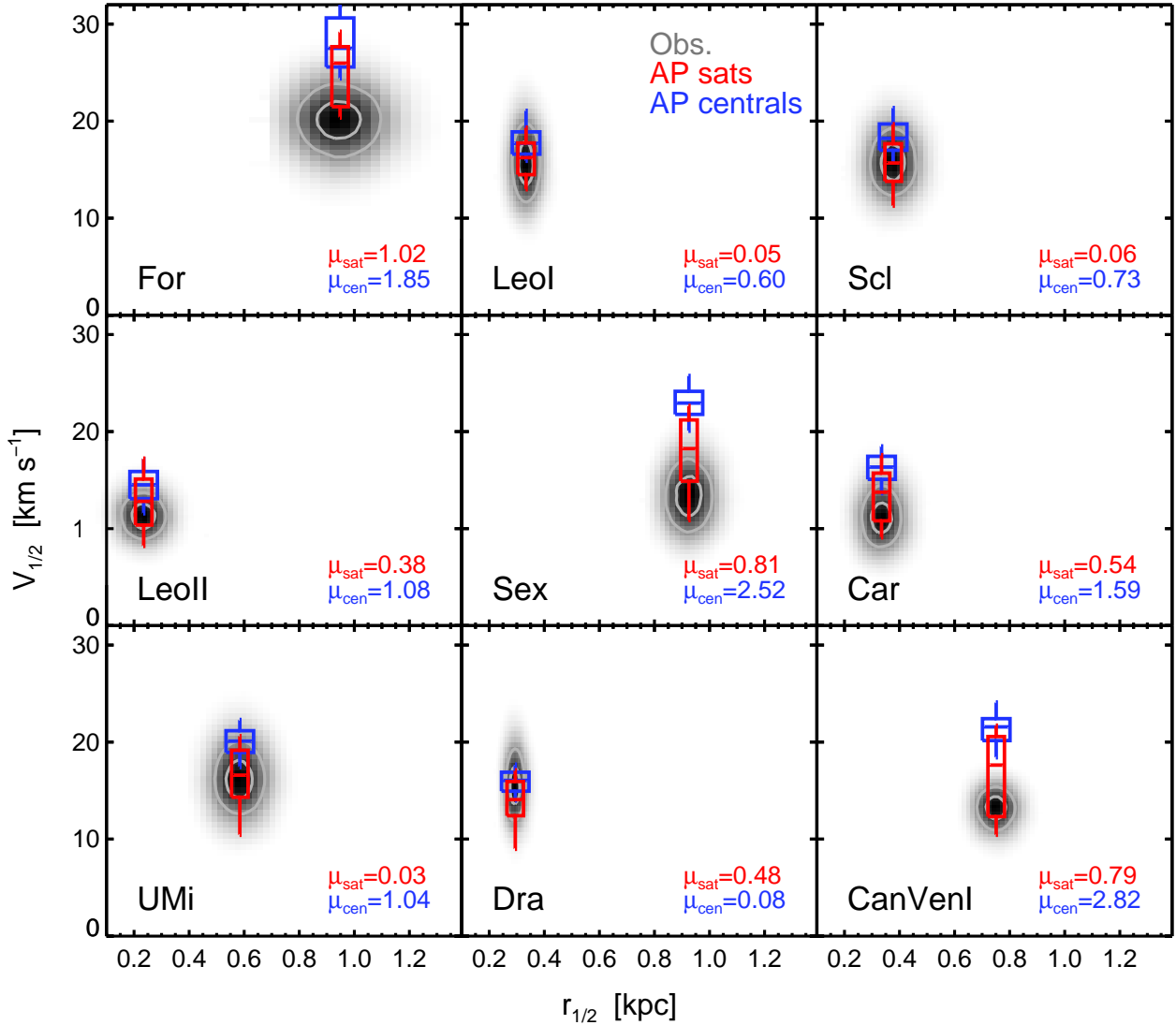}}\\%
  \caption{Circular velocity, $V_{1/2}$, of APOSTLE satellites
    matching the stellar mass of each of the $9$ Galactic dSphs, and
    measured at the observed half-light radius of each
    system. Observational estimates and uncertainties are given by the
    grey cloud, whereas red bar-and-whisker symbols indicate the
    values for matching APOSTLE satellites. Contours indicate the
    regions containing $50$ per cent and $80$ per cent of the
    distributions. Bars and whiskers represent the interquartile and
    $10$--$90^{\rm th}$ percentile intervals of predicted $V_{1/2}$,
    plotted at the median value of $r_{1/2}$.  Note the significant
    overlap between the satellite simulation results and the
    observational estimates; this may be quantified by the velocity
    difference between the mean observed and simulated values, divided
    by the combined rms of each distribution ($\mu$), which is less
    than unity in all cases. The values of $V_{1/2}$ for APOSTLE
    centrals are larger, since centrals have not experienced tidal
    stripping.}
\label{FigV}
\end{figure*}

\subsection{The dark matter content of APOSTLE satellites}
\label{SecDM}

Our main conclusion from Fig.~\ref{FigMstarM200} is that APOSTLE
satellites of given stellar mass are significantly less massive than expected
from abundance matching and, because of stripping, span a relatively
wide range of maximum circular velocities. Are these results
consistent with the observational constraints discussed in
Sec.~\ref{SecObs}? In other words, are the predicted values of
$r_{1/2}$ and $V_{1/2}$ consistent with those of Galactic
satellites of matching stellar mass? 

The main issue to consider when addressing this question is that the
half-light radii, $r_{1/2}$, of the faintest dSphs are smaller than
the smallest well-resolved radius in APOSTLE. This impacts the
analysis in two ways: one is that the faintest simulated dwarfs have
radii larger than observed\footnote{The subgrid equation of state
  imposed on star-forming gas particles by the EAGLE model results in
  a minimum effective radius of $\sim 400$ pc for galaxies in AP-L1
  runs \citep[see, e.g.,][]{Crain2015,Campbell2016}.}; another is that
the total mass enclosed by simulated dwarfs within radii as small as
the observed half-light radii might be systematically affected by the
limited resolution.

We illustrate this in the left panel of Fig.~\ref{FigSim} for the case
of the Sculptor dSph. The vertical shaded band shows the half-light
radius of that galaxy, $r_{1/2}=377^{+77}_{-73} \pc$ ($10$--$90^{\rm
  th}$ percentile interval), whereas the small crosses indicate the
stellar half-mass radii of Sculptor-like APOSTLE satellites on their
circular velocity profiles, $V_{\rm circ}(r)$. Clearly, for the
comparison with Sculptor to be meaningful, we should estimate masses
within the {\it observed} radius (grey band), rather than at the
half-mass radius of each of the simulated systems.

\begin{figure*}
  \hspace{-0.2cm}
  \resizebox{17cm}{!}{\includegraphics{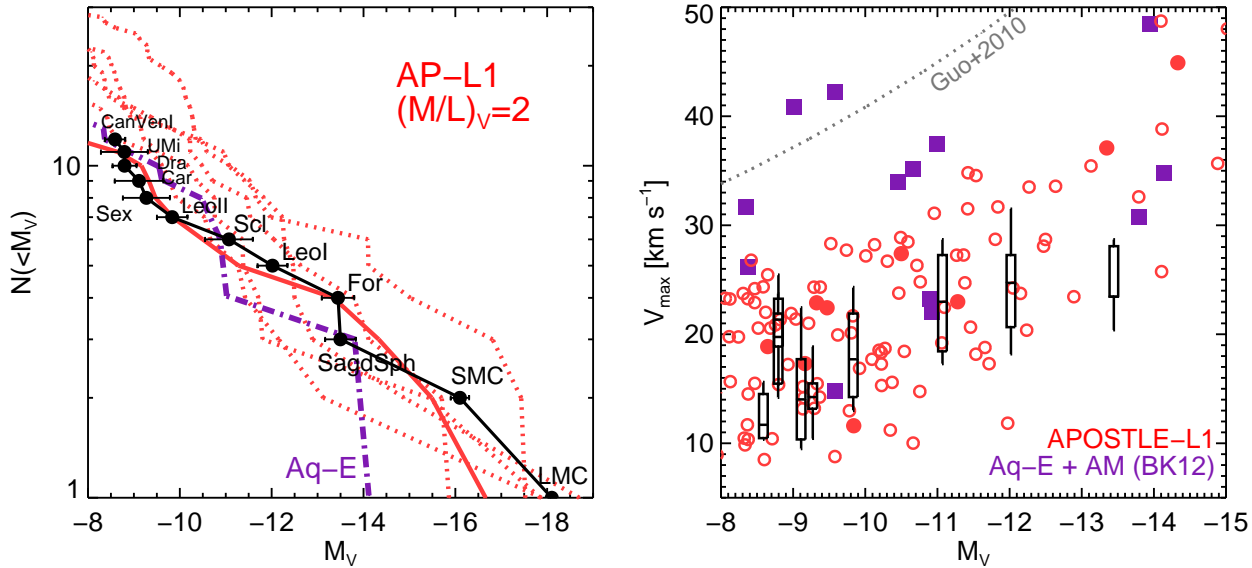}}\\%
  \caption{ {\it Left}: Luminosity function of satellites of the six
    APOSTLE L1 primaries (dotted lines), compared with that of
    Galactic satellites (filled circles).  We consider as satellites
    all systems within $300\kpc$ of the centre of each primary
    galaxy. We also show the luminosity function of the Aq-E halo
    (dot-dashed line) derived by \citet{Boylan-Kolchin2012} using an
    abundance matching model. The best fitting APOSTLE galaxy is
    highlighted with a solid line type. {\it Right}: $V_{\rm
      max}$--$M_V$ relation for APOSTLE satellites (circles) compared
    with the abundance matching estimates for Aq-E subhaloes from
    \citet{Boylan-Kolchin2012}.  Box-and-whisker symbols indicate the
    $V_{\rm max}$ range of APOSTLE satellites that match the stellar
    mass and $V_{1/2}$ of each of the 9 Galactic dSphs (see
    Sec. \ref{SecTBTF} for details). APOSTLE satellites inhabit
    markedly lower mass haloes than expected from abundance matching,
    at a given luminosity. APOSTLE also differs from Aq-E in the
    number of massive substructures. On average, each APOSTLE primary
    has $\sim 7.2$ satellite more massive than $V_{\rm max}>25\kms$;
    this number is actually only $5$ for the halo that best matches
    the MW satellite luminosity function (solid circles; two of them
    are brighter than $M_V=-15$). Aq-E has $21$ satellites with
    $V_{\rm max}$ exceeding $25\kms$, some of them fainter than
    $M_V=-8$. }
\label{FigLF}
\end{figure*}

However, the observed $r_{1/2}$ (although significantly larger than
the gravitational softening, which is fixed at $134$ pc at z=0 in AP-L1 runs)
is smaller the minimum resolved radius according to the convergence
criterion proposed by \citet{Power2003}. This is shown by the circular
velocity profiles in Fig.~\ref{FigSim}, where the line types change
from solid to dotted at the convergence radius, $r_{\rm conv}$
(defined by setting $\kappa=0.6$ in eq.~\ref{EqKappa}). The total
mass within $r_{1/2}$ and, consequently, $V_{1/2}$, are therefore
probably underestimated in the simulations. Fortunately, the analysis
of \citet{Power2003} also shows that mass profiles inside
$r_{\rm conv}$ deviate from convergence in a predictable fashion, so
that a correction procedure is straightforward to devise and
implement, at least for radii not too far inside $r_{\rm conv}$.

In Appendix~\ref{SecAppendix}, we describe the correction used to
estimate the total mass enclosed within the Sculptor half-light radius
for all Sculptor-like satellites in APOSTLE. The right-hand panel of
Fig.~\ref{FigSim} shows the APOSTLE $V_{1/2}$ estimates with and
without correction. These estimates are obtained by randomly sampling
the allowed range in $r_{1/2}$ as well as the $V_{\rm circ}(r_{1/2})$
distribution of Sculptor-like APOSTLE satellites. In brief, this
procedure involves: (i) choosing a random value for $r_{1/2}$
consistent with propagating the Gaussian errors in distance modulus
and angular size (see Sec.~\ref{SecObs} and col. 7 of
Table~\ref{TabSat}); (ii) measuring $V_{1/2}=V_{\rm circ}(r_{1/2})$
for a random satellite in AP-L1; and (iii) weighting\footnote{The
  weighting function is $\exp(-x^2/2\sigma^2)$, where
  $x=M_{\rm str}^{\rm AP}-M_{\rm str}^{\rm Scl}$, and $\sigma$ is the
  uncertainty in Sculptor's stellar mass discussed in
  Sec.~\ref{SecObs}.} each APOSTLE satellite by how closely it matches
Sculptor's stellar mass. (Although the procedure considers all
satellites, in practice only Sculptor-like satellites contribute
meaningfully, given the weighting procedure.)

The procedure is repeated $10,000$ times to derive the $V_{1/2}$
distribution shown in Fig.~\ref{FigSim}, which is then corrected for
resolution as described in Appendix~\ref{SecAppendix}. In the case of
Sculptor the correction to the measured $V_{1/2}$ values is relatively
mild; the median $V_{1/2}$ shifts only slightly, from $13.9 \kms$
before correction to $15.7 \kms$. This is actually the case for the
majority of systems; the largest correction is obtained for the Leo II
dSph, where the median $V_{1/2}$ increases by $24$ per cent, from
$10.3 \kms$ to $12.8 \kms$. Satellites like Fornax, which have larger
half-light radii well-resolved by APOSTLE, are corrected by less than
$5$ per cent.

We apply the same procedure outlined above to all $9$ Galactic
`classical' dSph satellites (excluding Sagittarius) and compare our
results with observational constraints in Fig.~\ref{FigV}.  The grey
shaded regions and contours denote the observational estimates
including uncertainties, while the red box-and-whisker symbols
indicate the results for APOSTLE satellites.  There is clearly
substantial overlap between observational estimates of $V_{1/2}$ and
the APOSTLE results for all $9$ dSphs, with no exception.

The values of $\mu_{\rm sat}$ quoted in the legends of Fig.~\ref{FigV}
indicate the absolute value of the difference between the mean
observed and simulated values, in units of the combined rms: the
difference is clearly not significant (less than unity) in any of the
$9$ cases. We conclude that the dark matter content of APOSTLE
satellites is in good agreement with the observed values.  We
emphasize that this agreement is {\it not} the result of cored DM
density profiles, as dwarf galaxies in APOSTLE show no evidence for cores
\citep{Oman2015,Sawala2016}.

We may assess the effect of tidal stripping on our conclusion by
repeating the above procedure using APOSTLE centrals, rather than
satellites, for the comparison. The results are shown by the blue
box-and-whisker symbols in Fig.~\ref{FigV} (red and blue boxes are
plotted with different widths, for clarity). The values of $V_{1/2}$
are systematically larger for centrals, since they have not
experienced tidal stripping.  The agreement is clearly poorer, in
particular for satellites `unusually large for their luminosities'
(Sec.~\ref{SecTS}) like Can~Ven~I, Sextans, Carina, and Fornax.

Consistency between APOSTLE and Galactic satellites therefore requires
that the dark matter content of at least some dSphs has been affected
by tides from the Milky Way halo. We emphasize that {\it all} subhaloes
have been affected by tides; their effects, however, are noticeable
mainly in systems whose sizes are large enough for their kinematics to
probe regions where the mass loss is significant. The strong
dependence of the effect of tides on galaxy size must be taken
carefully into account when comparing the dynamics of satellite and
isolated field galaxies to search for signs of environmental effects
\citep[see, e.g.,][]{Kirby2014}.

\subsection{The too-big-to-fail problem revisited}
\label{SecTBTF}

The previous section demonstrates that there is no conflict between
the dark matter content of APOSTLE satellites and that of Galactic
dSphs. This does not {\it per se} solve the `too-big-to-fail'
problem laid out by \citet{Boylan-Kolchin2011,Boylan-Kolchin2012},
which asserts that there is an excess of massive subhaloes without a
luminous counterpart in Milky Way-sized haloes. Does this problem
persist in APOSTLE? 

We have examined this question earlier in \citet{Sawala2016}, but we
review those arguments here in light of the revised uncertainties in
the mass of the Galactic classical dwarf spheroidals discussed in
Sec~\ref{SecObs}. Fig.~\ref{FigLF} reproduces the argument given by
\citet[][]{Boylan-Kolchin2012}. The solid squares in the right-hand
panel of our Fig.~\ref{FigLF} are taken directly from their Fig.~6 and
show the maximum circular velocities of the $13$ most luminous
subhaloes in the Aq-E halo, selected because, according to an
abundance matching model patterned after \citet{Guo2010}, its number
of satellites brighter than $M_V=-8$ matches that of the Milky
Way. This is shown by the magenta dot-dashed line in the left-hand
panel of Fig. ~\ref{FigLF}. The offset between the Aq-E solid squares
and the \citet{Guo2010} prediction (dotted line on right-hand panel)
is mainly due to tidal stripping.

Our APOSTLE L1 simulations also reproduce well the MW satellite
luminosity function (see dotted lines in left-hand panel), but they
differ from the \citet{Boylan-Kolchin2012} analysis of Aq-E in two
respects. One is that our subhaloes are, on average, significantly
less massive, at a given $M_V$, than assumed for Aq-E. This is because
the APOSTLE stellar mass -- halo mass relation is offset from
abundance matching predictions (see Fig.~\ref{FigMstarM200}).

The box-and-whisker symbols in the right-hand panel of
Fig. ~\ref{FigLF} show the $V_{\rm max}$ values (Table~\ref{TabSim})
of APOSTLE satellites that best match the stellar mass {\it and}
$V_{1/2}$ of each Galactic dSph. The procedure for estimating
$V_{\rm max}$ is the same as that outlined in Sec.~\ref{SecDM} for
computing $V_{1/2}$ but, in addition, weights each simulated satellite
by how closely it matches the observed $V_{1/2}$. These new
$V_{\rm max}$ estimates complement and extend those reported by
\citet{Sawala2016}.

The second difference concerns the number of massive substructures:
Aq-E has $21$ subhaloes with $V_{\rm max}>25 \kms$ within $300$ kpc
from the centre, $10$ of which are more luminous than $M_V=-8$,
according to the model of \citet{Boylan-Kolchin2012}. On the other
hand, APOSTLE L1 primaries have, on average, just $7.2\pm 2.5$
subhaloes with $V_{\rm max}>25 \kms$ within the same volume.  Indeed,
the APOSTLE primary whose satellite population best matches the MW
satellite luminosity function (solid red curve in the left-panel of
Fig.~\ref{FigLF}) has only $5$ subhaloes this massive, as indicated by
the solid circles in the right-hand panel of the same figure.  (Two of
those host satellites brighter than $M_V=-15$.)

As discussed by \citet{Sawala2016}, the reason for the discrepancy is
twofold. (i) Subhalo masses are systematically lower in cosmological
hydrodynamical simulations because of the reduced growth brought about
by the early loss of baryons caused by cosmic reionization and
feedback. This reduces the $V_{\rm max}$ of all subhaloes by
$\sim 12$ per cent. (ii) Chance plays a role too, as Aq-E
seems particularly rich in massive substructures. The average number
of subhaloes with $V_{\rm max}>25\kms$ expected {\it within the virial
  radius} of a halo of virial mass $M_{200}=1.2\times10^{12}\Msun$ is
just $8.1$ \citep{Wang2012a}, compared with $18$ for Aq-E, a
$>3\sigma$ upward fluctuation. Note that
the expected number would drop to just $5.4$ after correcting for the
$\sim 12$ per cent reduction in $V_{\rm max}$. Indeed, the subhalo
velocity function is so steep that even a slight variation in
$V_{\rm max}$ leads to a disproportionately large change in the number
of massive substructures.

The discrepancy between APOSTLE and Aq-E noted above can therefore be
ascribed to a chance upward fluctuation in the number of massive
substructures in Aq-E coupled with the reduction of subhalo masses due
to the loss of baryons in a hydrodynamical simulation.

\begin{figure}
  \hspace{-0.2cm}
  \resizebox{8cm}{!}{\includegraphics{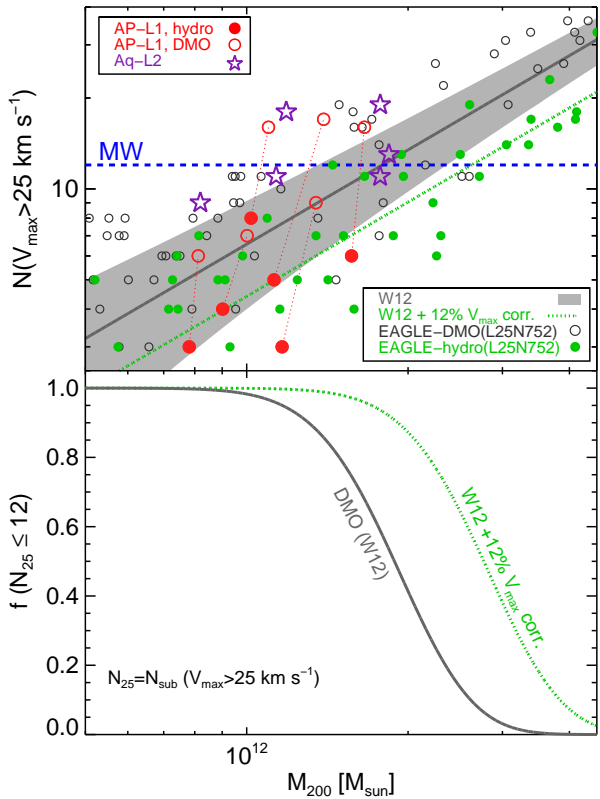}}\\%
  \caption{{\it Top}: Number of massive subhaloes
    ($V_{\rm max}>25\kms$) within $r_{200}$, shown as a function of 
    virial mass, $M_{200}$, for APOSTLE-L1 (solid red
    circles) haloes and their DMO counterparts (open red circles), and
    Aquarius haloes (open stars). The results from EAGLE L0250752-Ref
    and its DMO counterpart are shown using small solid green circles
    and grey open circles, respectively. The prediction of
    \citet[][W12]{Wang2012a} with $1\sigma$ scatter is shown by the
    grey band. Including the $12$ per cent reduction in $V_{\rm max}$
    brings the W12 relation down to the green dotted line.  The
    horizontal dashed line indicates the number of MW satellites
    brighter than $M_V=-8$.  {\it Bottom}: The fraction of haloes with
    $12$ or fewer massive subhaloes (i.e., $V_{\rm max}>25\kms$), as a
    function of the virial mass of the primary. The grey curve
    corresponds to the \citet{Wang2012a} estimate from
    dark-matter-only simulations. The green curve includes the $12$ per cent
    reduction in $V_{\rm max}$ obtained in hydrodynamical simulations. }
\label{FigNsat}
\end{figure}

\subsection{TBTF and the mass of the Milky Way}
\label{SecMWMass}

The number of massive substructures is, of course, quite sensitive to
the virial mass of the host halo. Following \citet{Wang2012a} and
\citet{Cautun2014}, we may use this to derive an upper
limit to the mass of the Milky Way. In APOSTLE every subhalo with
$V_{\rm max}>25 \kms$ hosts a satellite brighter than $M_V=-8$ (or,
equivalently, more massive than $M_{\rm str}\sim 10^5\Msun$; see
the right-hand panel of Fig.~\ref{FigMstarM200}). This means that any
potential Milky Way host halo with more than $\sim 12$ subhaloes this
massive will either suffer from a `too-big-to-fail' problem or have
an excess of luminous satellites. 

We examine this in Fig.~\ref{FigNsat}, which shows the number of
massive subhaloes within $r_{\rm 200}$ as a function of virial
mass. The criterion above implies that, in the top panel, only systems
below the dashed line labelled `MW' are likely to reproduce well the
MW satellite population. The grey band in the same panel shows the
expected number ($\pm 1\sigma$) of massive substructures according to
\citet{Wang2012a}. Small (grey) open circles indicate the results from
a $752^3$-particle dark-matter-only simulation of a cube $25$ Mpc on a
side. Small (green) filled circles correspond to the same volume, but
for a run including baryons and the full galaxy formation physics
modules from the EAGLE project\footnote{This simulation is labelled
  L0250752 in \citet{Schaye2015} and was run using the
    parameters of the `Ref' model.}. The offset between green and grey
circles demonstrates the effect of the reduction of $V_{\rm max}$
caused by the inclusion of baryons in the simulation.

The six Aquarius haloes \citep{Springel2008b} are shown by starred
symbols: these systems are slightly overabundant in massive
substructures relative to both the EAGLE runs and the predictions of
\citet{Wang2012a}, which are based on large samples of haloes from the
Millennium Simulations. Three Aquarius haloes have more
than 12 massive substructures, and therefore would not be
consistent with the MW satellite population according to our APOSTLE
results.
  
Including baryons changes this, as shown by the six primaries in
APOSTLE L1: these are shown in Fig.~\ref{FigNsat} with red circles;
filled symbols for the hydrodynamical runs, and open symbols for the DMO
versions. The DMO runs give results similar to Aquarius: half of
APOSTLE DMO are above the `MW' line. The number of massive
substructures drops substantially once baryons are included (filled
circles), so that all six primaries in the APOSTLE L1
hydrodynamical runs are actually consistent with the MW.

We may use these results to derive firm upper limits on the mass of
the Milky Way. This is shown in the bottom panel of Fig.~\ref{FigNsat}
where each curve traces the fraction of haloes of a given virial mass
that have $12$ (or fewer) massive substructures (i.e., the observed
number of Galactic satellites brighter than $M_V=-8$). We show results
for two cases; one where the numbers are derived from the formula of
\citet{Wang2012a}, assuming Poisson statistics (solid grey lines) and
another where the zero-point of that relation has been shifted to
account for the $12$ per cent reduction in $V_{\rm max}$ discussed above (see
green dotted line in the top panel of Fig.~\ref{FigNsat}).

Clearly, the reduction in $V_{\rm max}$ induced by the loss of baryons
in hydrodynamical simulations significantly relaxes the constraints
based on dark-matter-only simulations. Indeed, according to this
argument, fewer than $5$ per cent of haloes more massive than
$2.8\times 10^{12}\Msun$ can host the Milky Way, assuming the DMO
relation. The same criterion results in an increased mass limit of
$4.2\times 10^{12}\Msun$ adopting the $V_{\rm max}$ correction. This
may also be compared with the earlier analysis of \citet{Wang2012a},
which found an upper limit of $2\times 10^{12}\Msun$, and of
\citet{Cautun2014}, which derived an even stricter limit, albeit using
slightly different criteria.

\section{Summary and Conclusions}
\label{SecConc}

We use the APOSTLE suite of $\Lambda$CDM cosmological hydrodynamical
simulations of the Local Group to examine the masses of satellite
galaxies brighter than $M_V=-8$ (i.e., $M_{\rm str}>10^5 \Msun$).  Our
analysis extends that of \citet{Sawala2016}, were we showed that our
simulations reproduce the Galactic satellite luminosity function
and show no sign of either the `missing satellites' problem nor of the
`too-big-to-fail' problem highlighted in earlier work. Our main
conclusions may be summarised as follows.

Previous studies have underestimated the uncertainty in the mass
enclosed within the half-light radii of Galactic dSphs, derived from
their line-of-sight velocity dispersion and half-light radii. Our
analysis takes into account the error propagation due to uncertainties
in the distance, effective radius, and velocity dispersion, and also
include an estimate of the intrinsic dispersion of the modeling
procedure, following the recent work of \citet{Campbell2016}. The
latter is important as it introduces a base systematic uncertainty
that exceeds $\sim 20$ per cent.

Simulated galaxies in APOSTLE/EAGLE follow a stellar mass -- halo mass
relation that differs, for dwarf galaxies, from common extrapolations
of abundance matching models, a difference that is even more
pronounced for satellites due to tidal stripping, At fixed stellar
mass, APOSTLE dwarfs inhabit halos significantly less massive than AM
predicts. This difference, however, might not be readily apparent
because tides strip halos from the outside in and some dSphs are too
compact for tidal effects to be readily apparent.

We find that the dynamical mass of {\it all} Galactic dSphs is in
excellent agreement with that of APOSTLE satellites that match their
stellar mass. APOSTLE centrals (i.e., not satellites), on the
other hand, overestimate the observed mass of four Galactic dSphs (Can
Ven~I, Sextans, Carina, and Fornax), suggesting that they have had
their dark matter content significantly reduced by stripping. The
other, more compact, dSphs are well fit by either APOSTLE satellites
or centrals, so tides are not needed to explain their dark matter
content.

After accounting for tidal mass losses, we find that all APOSTLE halos
(satellites or centrals) with $V_{\rm max}>25 \kms$ host dwarfs
brighter than $M_V=-8$. Only systems with fewer than $\sim 12$
subhaloes with $V_{\rm max}>25$ km/s are thus compatible with the
population of luminous MW satellites. This suggests an upper
limit to the mass of the Milky Way halo: we find that most halos with
virial mass not exceeding $2\times 10^{12}\, M_\odot$ should pass this
test, unless they are unusually overabundant in massive substructures.

Our APOSTLE primaries satisfy these constraints, and show a dwarf
galaxy population in agreement with observations of the
Local Group, including their abundance as a function of mass, their
dark matter content, and their global kinematics. Furthermore, APOSTLE
uses the same galaxy formation model that was found by EAGLE to
reproduce the galaxy stellar mass function in cosmologically
significant volumes. We consider this a significant success for direct
simulations of galaxy formation based on the $\Lambda$CDM paradigm.

We note that this success does not require any substantial
modification to well-established properties of $\Lambda$CDM. In
particular, none of our simulated dwarf galaxies have `cores' in
their dark mass profiles, but yet have no trouble reproducing the
detailed properties of Galactic satellites. Baryon-induced cores are
not mandatory to solve the `too-big-to-fail' problem.

We end by noting that a number of recent studies have argued that
TBTF-like problems also arise when considering the properties of M31
satellites \citep{Tollerud2014,Collins2014}, as well as those of field
galaxies in the local Universe
\citep{Garrison-Kimmel2014,Papastergis2015}. It remains to be seen
whether the resolution we advocate here for Galactic satellites will
solve those problems as well. We plan to report on those issues in
future work.

\section{Acknowledgements}

We acknowledge useful discussions with Mike Boylan-Kolchin, Manolis
Papastergis, and Alan McConnachie. The research was supported in part
by the Science and Technology Facilities Council Consolidated Grant
(ST/F001166/1), and the European Research Council under the European
Union’s Seventh Framework Programme (FP7/2007-2013)/ERC Grant
agreement 278594-GasAroundGalaxies. CSF acknowledges ERC Advanced
Grant 267291 ’COSMIWAY’; and JW the 973 program grant 2015CB857005 and
NSFC grant No. 11373029, 11390372. This work used the DiRAC Data
Centric system at Durham University, operated by the Institute for
Computational Cosmology on behalf of the STFC DiRAC HPC Facility
(www.dirac.ac.uk), and also resources provided by WestGrid
(www.westgrid.ca) and Compute Canada (www.computecanada.ca). The DiRAC
system was funded by BIS National E-infrastructure capital grant
ST/K00042X/1, STFC capital grants ST/H008519/1 and ST/K00087X/1, STFC
DiRAC Operations grant ST/K003267/1 and Durham University. DiRAC is
part of the National E-Infrastructure. This research has made use of
NASA's Astrophysics Data System.

\bibliographystyle{apj}
\bibliography{master}
\bsp

\appendix

\section{Numerical Corrections}
\label{SecAppendix}

\begin{figure*}
  \hspace{-0.2cm}
  \resizebox{17cm}{!}{\includegraphics{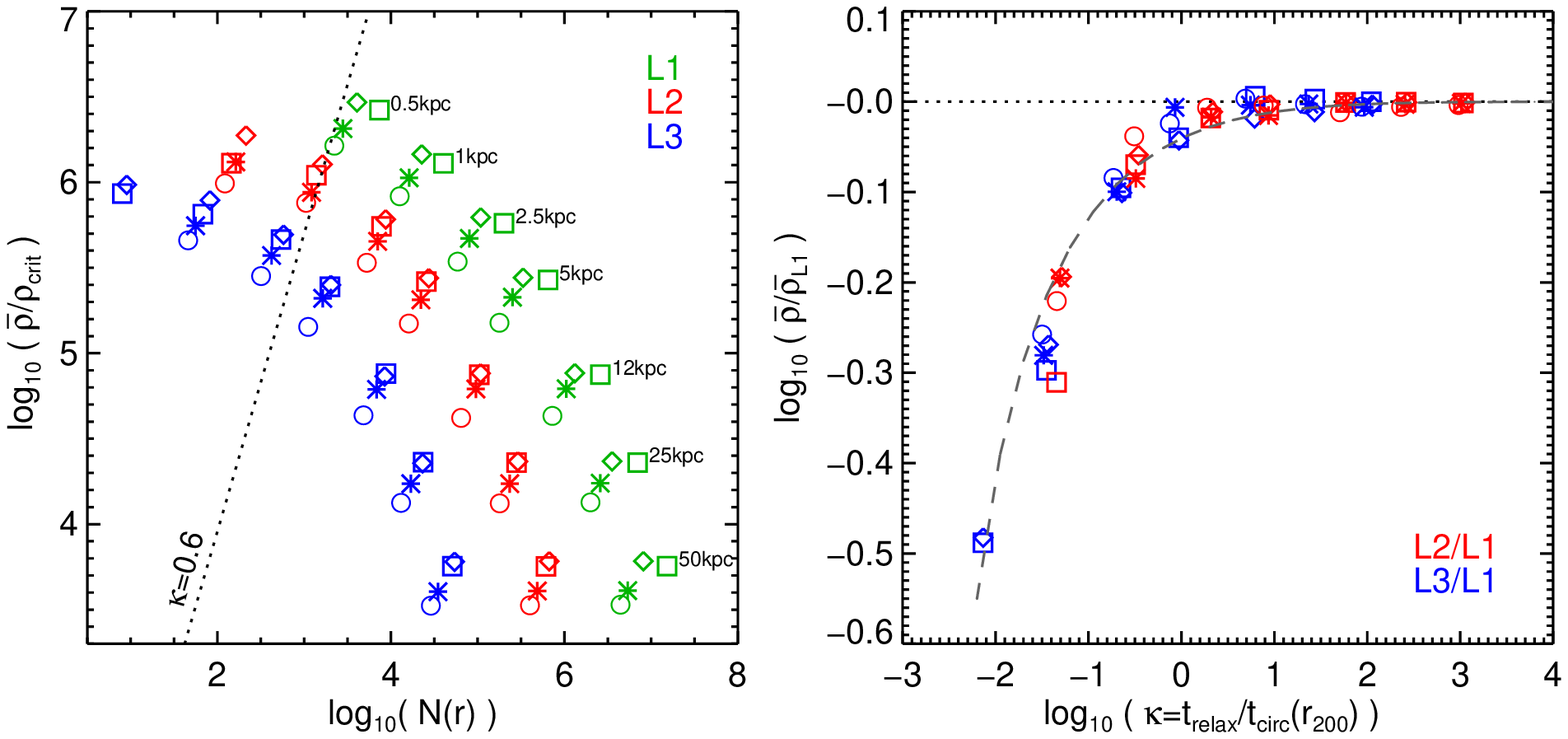}}\\%
  \caption{{\it Left}: Mean inner density as a function of enclosed
    number of particles at different radii, as labelled, for $4$
    primary haloes in dark-matter-only APOSTLE, simulated at three
    different resolution levels (L1 to L3). Different haloes are shown
    by different symbols, and colours indicate different resolution
    levels. Points to the right of the $\kappa=0.6$ line
    (eq.~\ref{EqKappa}) converge in circular velocity to better than
    $15$ per cent \citep{Power2003}. Two L3 haloes have fewer than 5
    particles at the smallest radius, and are not shown. {\it Right}:
    Mean inner density as a function of $\kappa$ for L2 and L3 haloes,
    normalized to the values obtained for the highest-resolution run,
    L1. Symbols are the same as in the left panel. The dashed line has
    been used to correct densities at small radii.  }
\label{FigP03}
\end{figure*}

As discussed by \citet{Power2003}, the enclosed mass profiles of N-body
realizations of $\Lambda$CDM haloes converge
outside a minimum radius, $r_{\rm conv}$, that depends on the number of particles
enclosed and on the mean inner density of the halo at that
radius. This is because simulated profiles converge only for radii
where the two-body relaxation timescale is long compared with the age
of the Universe. A criterion for convergence may thus be derived using
the ratio of relaxation time to the circular velocity at the virial radius:

\begin{equation}
 \kappa(r)= \frac{t_{\rm relax}(r)}{t_{\rm circ}(r_{200})}=
  \frac{N(r)}{8\,\ln\,N(r)}\frac{r/V_c}{r_{200}/V_{200}},
\end{equation}
which may also be written as,
\begin{equation}
 \kappa(r)= \frac{\sqrt{200}}{8}\frac{N(r)}{\ln\,N(r)}\bigg
  (\frac{\bar{\rho}(r)}{\rho_{\rm crit}}\bigg )^{-1/2}.
\label{EqKappa}
\end{equation}
where $N(r)$ is the enclosed number of particles and $\bar{\rho}(r)$ is the
mean density inside the radius $r$. At radii where $\kappa \approx 1$ profiles
converge to better than $10$ per cent in terms of circular velocity. Stricter
convergence demands larger values of $\kappa$ and implies, consequently,
larger values of $r_{\rm conv}$ \citep{Navarro2010}.

Fig.~\ref{FigP03} illustrates this for the case of the
dark-matter-only realizations of four different APOSTLE primary haloes, run
at three different resolutions, each differing by about a factor of
$\sim 10$ in particle mass and $\sim 2$ in force resolution (L1 to L3,
where L1 is best resolved). 

The left panel of Fig.~\ref{FigP03} shows the mean enclosed density at
various radii. Different colours indicate the various resolution
levels, whereas different symbols correspond to different haloes. At
large radii all resolutions converge to the same result (i.e., like
symbols line up horizontally). At small radii, however, the
lower-resolution profiles gradually deviate from the
highest-resolution (L1) run. The \citet{Power2003} criterion is shown
by the inclined dotted line, for $\kappa=0.6$. Note that points
clearly converge, regardless of resolution, to the right of this line,
but those on the left deviate noticeably from the results obtained for
the highest-resolution case, L1.

The smooth trend in density contrast with enclosed particle number
suggests a simple way to correct an under-resolved halo
profile. Indeed, expressed in terms of $\kappa$, the `deficit' in
density observed in the inner regions always follows the same
pattern. This is shown in the right-hand panel of Fig.~\ref{FigP03},
where we show, for all radii $\geq 0.5\kpc$ and all haloes, the
density in units of the `true' values obtained for L1. All haloes
follow the same pattern, which we approximate with a fitting function,
$\log (1-{\bar \rho/{\bar \rho_{\rm conv}}}) = a (\log \kappa)^2 + b
(\log \kappa) + c$ where (a,b,c)=($-0.04$,$-0.5$,$-1.05$).

Since densities of L1 are `converged' according to the left-hand panel
of Fig.~\ref{FigP03} (${\bar \rho_{\rm L1}}={\bar \rho_{\rm conv}}$),
the aforementioned trend may be used to extrapolate the results of a
simulation to radii smaller than the traditional value of $r_{\rm
  conv}$ dictated by assuming $\kappa=0.6$.

\begin{figure*}
  \hspace{-0.2cm}
  \resizebox{17cm}{!}{\includegraphics{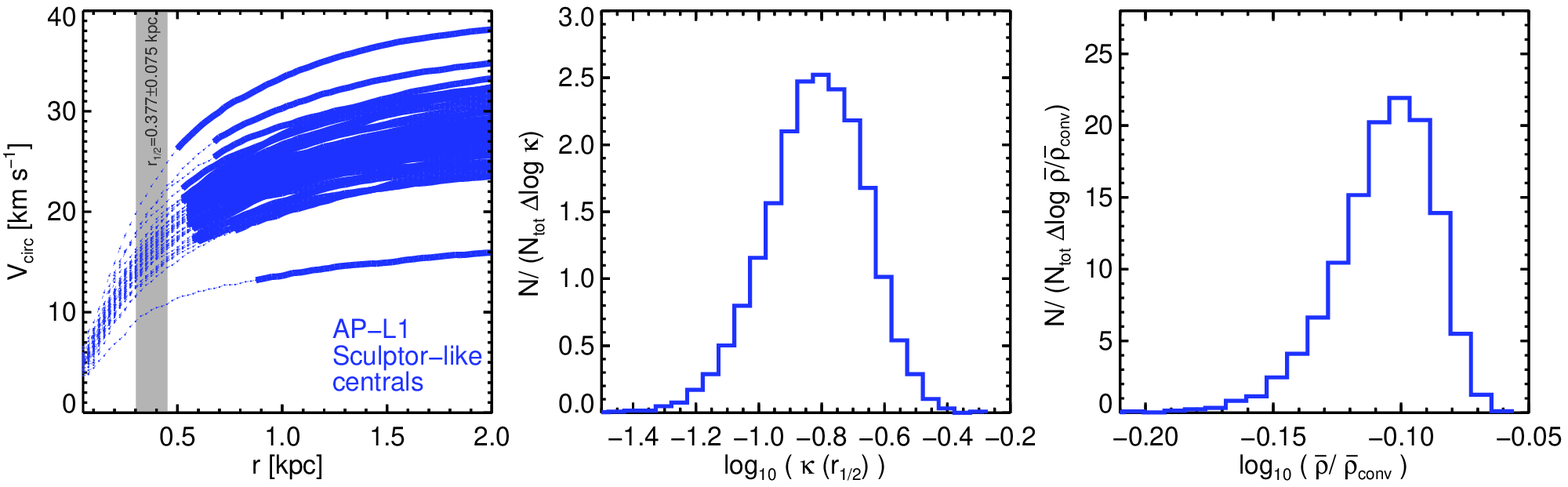}}\\%
  \caption{{\it Left}: Circular velocity curves of Sculptor-like
    APOSTLE centrals. The grey band corresponds to the $10$--$90^{\rm
      th}$ percentile interval for the observed $r_{1/2}$ of
    Sculptor. {\it Middle}: Distribution of $\kappa(r)=t_{\rm
      relax}(r)/t_{\rm circ}(r_{200})$ for the rotation curves in the
    left panel, at radii consistent with the $r_{1/2}$ of
    Sculptor. {\it Right:} Distribution of the density correction
    factor derived from the $\kappa$ distribution of the middle panel
    and the fit presented in the right panel of Fig.~\ref{FigP03}.}
\label{FigA2}
\end{figure*}

We show in Fig.~\ref{FigA2} the results of applying this correction to
Sculptor-like central galaxies in APOSTLE-L1. Typical values of
$\kappa$ at $r_{1/2}$ for this galaxy are about $0.15$, which results
in a correction in enclosed density of roughly $20$ per cent. The
distribution shown in the right-hand panel of Fig.~\ref{FigA2} is then
used statistically to correct the raw $V_{1/2}$ estimates from our L1
satellites. The result of applying this to Sculptor-like satellites is
shown in Fig.\ref{FigSim}. This same procedure is applied separately
to each Galactic satellite in order to derive the predictions shown in
Fig.~\ref{FigV}.

\begin{table*}
  \caption{The parameters of classical dSph satellites of MW.}
  \bc
  \def\arraystretch{2.0}
  \setlength{\tabcolsep}{5pt} 
  \begin {tabular*}{18.cm}{{l} *{4}{r} {c} *{4}{l} }
    \hline
    name   &   $(m-M)_0$   & $R_{\rm eff}$~~~~ &  $\sigma_{\rm los}~~~$ & $m_V~~~~$& $\frac{M_{\rm str}}{L_V}$ & 3D $r_{1/2}$  &  $M_{1/2}$       &  $V_{1/2}$  & $M_{\rm str}$  \\
           &              &   (arcmin)  &  ($\kms$)     &          &                &    (\pc)     &  ($10^7 \Msun$) & ($\kms$)   & ($10^6\Msun$)     \\ 
    ref.   &    (a)~~~~   &             &              &   (a)~~~~  &      (b)         &              \\
    \hline

    For  &  $20.84\pm0.18$ & $16.6\pm1.2^{a,c}$& $11.7\pm0.9^f$& $7.4\pm0.3$ & $1.2$ & $950^{+70(140)}_{-70(130)}$& $8.9^{+1.9(3.8)}_{-1.7(3.0)}$   & 
    $20.1\pm1.9(3.6)$ & $24^{+6(13)}_{-5(9)}$\\
    Leo I   &  $22.02\pm0.13$ & $ 3.4\pm0.3^{a,c}$& $9.2\pm1.4^g$ & $10.0\pm0.3$& $0.9$ & $334^{+24(47)}_{-24(44)}$  & $1.9^{+0.6(1.2)}_{-0.6(0.8)}$ &
    $15.8\pm2.0(3.9)$ & $5.0^{+1.1(2.4)}_{-1.0(1.7)}$\\
    Scl&  $19.67\pm0.14$ & $11.3\pm1.6^{a,c}$& $9.2\pm1.1^f$ & $8.6\pm0.5$ & $1.7$ & $377^{+40(77)}_{-39(73)}$  & $2.2^{+0.6(1.2)}_{-0.5(0.9)}$ & 
    $15.8\pm1.8(3.4)$ & $3.9^{+1.5(3.4)}_{-1.1(1.8)}$\\
    Leo II& $21.84\pm0.13$ & $2.6\pm0.6^{a,c}$ & $6.6\pm0.7^h$ & $12\pm0.3$ & $1.6$ & $235^{+38(73)}_{-38(71)}$ & $0.68^{+0.21(0.4)}_{-0.17(0.3)}$ & $11.3\pm1.2(2.3)$ & $1.2^{+0.3(0.6)}_{-0.2(0.4)}$ \\ 
    Sex I&$19.67\pm0.10$ & $27.8\pm1.2^{a,c}$& $7.9\pm1.3^f$ & $10.4\pm0.5$& $1.6$ & $926^{+40(77)}_{-39(73)}$  & $3.9^{+1.2(2.5)}_{-1.0(1.7)}$ & 
    $13.5\pm1.9(3.5)$ & $0.7^{+0.3(0.6)}_{-0.2(0.3)}$\\
    Car    &$20.11\pm0.13$ & $ 8.2\pm1.2^{a,c}$& $6.6\pm1.2^f$ & $11.0\pm0.5$& $1.0$ & $334^{+36(69)}_{-35(66)}$  & $1.0^{+0.3(0.7)}_{-0.3(0.5)}$ & 
    $11.3\pm1.7(3.1)$ & $0.38^{+0.14(0.32)}_{-0.11(0.18)}$\\
    UMi&$19.40\pm0.10$ & $19.9\pm1.9^d$& $9.5\pm1.2^i$ & $10.6\pm0.5$& $1.9$ & $584^{+42(82)}_{-41(78)}$  & $3.6^{+1.0(2.0)}_{-0.8(1.4)}$ &
    $16.3\pm1.9(3.6)$ & $0.54^{+0.21(0.46)}_{-0.15(0.25)}$\\
    Dra     &$19.40\pm0.17$ & $10.0^{+0.3, e}_{-0.2}$& $9.1\pm1.2^j$ & $10.6\pm0.2$& $1.8$ & $294^{+17(33)}_{-16(30)}$  & $1.7^{+0.5(0.9)}_{-0.4(0.7)}$ &
    $15.6\pm1.9(3.6)$ & $0.51^{+0.09(0.20)}_{-0.09(0.15)}$ \\
    CVn I& $21.69\pm0.10$&$8.9\pm0.4^e$& $7.6\pm0.4^k$& $13.1\pm0.2$& $1.6$& $751^{+34(64)}_{-32(60)}$& $3.0^{+0.5(1.1)}_{-0.5(0.9)}$ &
    $13.1\pm1.1(2.1)$ & $0.37^{+0.06(0.13)}_{-0.05(0.10)}$\\ 
    \hline
  \end{tabular*}
  \ec
  \begin{tablenotes}
  \item \textbf{Notes}: Uncertainties in the observed parameters are
    taken directly from the references. We assume in all cases that
    they correspond to standard deviations of a Gaussian error
    distribution. The uncertainties quoted for derived parameters,
    i.e. the last four columns, correspond to interquartile and
    $10$--$90^{\rm th}$ percentile intervals, written outside and
    inside parentheses, respectively.
    \item \textbf{References}: $^a$\citet{McConnachie2012};
      $^b$\citet{Woo2008}; $^c$\citet{Irwin1995};
      $^d$\citet{Palma2003}; $^e$\citet{Martin2008};
      $^f$\citet{Walker2009c}; $^g$\citet{Mateo2008};
      $^h$\citet{Koch2007}; $^i$\citet{Walker2009d};
      $^j$\citet{Walker2007}; $^k$\citet{Simon2007}.
      
  \end{tablenotes} 
  \label{TabSat}
\end{table*}

\begin{table*}
  \caption{Parameters of APOSTLE satellites matching the stellar mass
    of Galactic classical dSph satellites.}
  \bc
  \def\arraystretch{2.0}
  \begin {tabular*}{10cm}{{l}  *{3}{l} }
    \hline
                          &    $M_{\rm 1/2}$   & $V_{\rm 1/2}$  &   $V_{\rm max}$   \\  
                           &      ($10^7 \Msun$)     &  ($\kms$)           &   ($\kms$)        \\
    \hline
    Fornax-like (14)        &    $13^{+3(6)}_{-3(5)}$  &  $25.5^{+1.8(3.4)}_{-4.0(5.1)}$  &  $23.0^{+4.6(4.6)}_{-0(3.07)}$   \\
    Leo I-like (37)         &    $2.0^{+0.5(1.0)}_{-0.5(0.8)}$  &  $16.2^{+1.5(3.1)}_{-1.7(3.3)}$  &  $24.7^{+2.6(6.7)}_{-4.1(6.5)}$   \\
    Sculptor-like(56)      &    $2.1^{+0.8(1.6)}_{-0.6(1.2)}$  &  $15.7^{+2.0(4.0)}_{-1.9(4.4)}$  &  $23.0^{+4.2(5.5)}_{-4.6(5.7)}$  \\
    Leo II-like(50)       &    $0.9^{+0.5(1.0)}_{-0.4(0.6)}$  &  $12.8^{+2.3(4.4)}_{-2.4(4.6)}$  &  $17.7^{+4.0(6.6)}_{-3.5(4.5)}$  \\
    Sextans-like(89)       &    $7.1^{+2.7(4.2)}_{-2.3(4.6)}$  &  $18.2^{+3.0(4.4)}_{-3.4(7.4)}$  &  $14.2^{+1.2(4.1)}_{-1.1(3.8)}$   \\
    Carina-like(117)      &    $1.4^{+0.6(1.2)}_{-0.5(0.8)}$  &  $13.8^{+2.0(3.9)}_{-3.0(4.7)}$   &  $14.0^{+3.2(8.0)}_{-3.7(4.5)}$   \\
    Ursa Minor-like(95)   &    $3.7^{+1.2(2.4)}_{-1.1(2.3)}$  &  $16.6^{+2.6(4.1)}_{-4.2(5.7)}$  &  $19.8^{+2.2(4.0)}_{-4.3(5.6)}$   \\
    Draco-like(48)       &    $1.3^{+0.4(0.8)}_{-0.4(0.8)}$  &  $14.7^{+1.9(3.1)}_{-1.7(5.1)}$  &  $21.3^{+1.9(4.1)}_{-2.5(6.0)}$    \\
    Canes Venatici I-like(45)&  $5.3^{+2.0(3.0)}_{-2.6(3.4)}$   &  $17.6^{+2.9(4.1)}_{-5.2(7.2)}$  &  $11.7^{+2.8(3.8)}_{-1.2(1.8)}$   \\
  \end{tabular*}
  \ec
  \begin{tablenotes}
  \item \textbf{Notes}: Values of $M_{1/2}$ and $V_{1/2}$ have been
    corrected by the procedure outlined in
    Appendix~\ref{SecAppendix}. $V_{\rm max}$ values are obtained by
    matching the stellar mass {\it and} $V_{1/2}$ of APOSTLE
    satellites to those of MW satellites. See text for details.
    Numbers quoted in parentheses after the names are the number of
    simulated satellites matching the stellar mass of the
    corresponding MW dSph. Similar to Table~\ref{TabSat},
    uncertainties represent interquartile and $10$--$90^{\rm th}$
    percentile intervals, written outside and inside parentheses,
    respectively.
  \end{tablenotes} 
  \label{TabSim}
\end{table*}

\label{lastpage}

\end{document}